\documentclass[aps,twocolumn,superscriptaddress,nofootinbib,pre,floatfix]{revtex4-1}

\usepackage{polski}

\usepackage{url}

\usepackage[pdftex]{graphicx}
\usepackage{latexsym,amsmath,verbatim}
\usepackage{xcolor}

\usepackage[hidelinks,unicode=true]{hyperref}
\hypersetup{colorlinks=true,linkcolor=blue,urlcolor=blue,citecolor=blue,pdfhighlight=/N
}
\usepackage{lipsum}
\usepackage[english]{babel}
\usepackage{microtype}

\usepackage{bbold}
\usepackage[normalem]{ulem}

\newcommand{\be}{\begin{equation}}
\newcommand{\ee}{\end{equation}}

\graphicspath{{./figures/}}
\begin{document}
\title{Limited efficacy of forward contact tracing in epidemics}

\author{Giulia de Meijere}
\affiliation{Gran Sasso Science Institute, Viale  F.  Crispi  7,  67100  L’Aquila, Italy}
\affiliation{Istituto dei Sistemi Complessi (ISC-CNR), Via dei Taurini 19,
I-00185 Roma, Italy}

\author{Claudio Castellano}
\affiliation{Istituto dei Sistemi Complessi (ISC-CNR), Via dei Taurini 19,
I-00185 Roma, Italy}
\affiliation{Centro Ricerche Enrico Fermi, Piazza del Viminale, 1,
  I-00184 Rome, Italy}

\begin{abstract}
  Infectious diseases that spread silently through asymptomatic or
  pre-symptomatic infections represent a challenge for policy makers.
  A traditional way of
  achieving isolation of silent infectors from the community is
  through forward contact tracing, aiming at identifying individuals
  that might have been infected by a known infected person.
  In this work we investigate how efficient this measure is
  in preventing a disease from becoming endemic.
  We introduce an SIS-based compartmental model where 
  symptomatic individuals may self-isolate and trigger a contact tracing
  process aimed at quarantining asymptomatic infected individuals.
  Imperfect adherence and delays affect both measures.
  We derive the epidemic threshold analytically and find
  that contact tracing alone can only lead to a very limited increase
  of the threshold.
  We quantify the effect of imperfect adherence and the impact of
  incentivizing asymptomatic and
  symptomatic populations to adhere to isolation. Our
  analytical results are confirmed by simulations on complex networks
  and by the numerical analysis of a much more complex model
  incorporating more realistic in-host disease progression.
\end{abstract}

\maketitle

\section*{Introduction}

Containing the propagation of infectious diseases that are able to
spread silently through asymptomatic or pre-symptomatic infections is
particularly challenging \cite{Fraser, Ferretti, Johansson, Li, Groendyke}.
A traditional way of achieving isolation of silent infectors without
applying restrictive policies to entire populations (such as lockdowns)
is through the so-called contact tracing (CT) measure, considered in the past to cope with outbreaks of SARS \cite{Riley2003, Lipsitch2003}, Foot and Mouth disease \cite{Ferguson2001}, smallpox \cite{Foege1971, Riley2006}, tubercolosis \cite{Fox2013}, HIV \cite{Rutherford1988}, Ebola \cite{Saurabh2017,Swanson2018}, SARS-CoV-2 \cite{Shin2020}.
Being at higher risk of infection, the contacts of a known infected
person are retrieved and recommended to
quarantine. However this measure still constitutes a considerable
social burden, as it may isolate also healthy individuals from the community and
requires intense logistical efforts for tracing contacts - a
particularly difficult task for airborne-diseases like COVID-19.
As fatigue sets in and adherence to isolation mandates and to
pharmaceutical interventions fades \cite{Silva2020}, it is not clear under what
conditions it is convenient to simplify the policies and rely only
on case-isolation and vaccination strategies or when, instead,
implementing CT is crucial for epidemic control.
This is the question we tackle in this work.

The success of contact tracing in the past has not been universal. While some outbreaks could be controlled \cite{Gibney2020}, others required more intense interventions to achieve epidemic control \cite{Blakely2020,Knapton2020}.
The efficacy of the contact tracing measure
has been studied in a number of works, with various approaches ranging
from stochastic simulations (\cite{Mancastroppa21, Kerr, Burdinski, Groendyke, Hellewelllancet, Hasegawa, Horstmeyer}) to analytical investigations
(\cite{Lee,Heidecke, House, ReynaLara, Bianconi, kojaku21, Burgio, Rizi}).

The utility of contact tracing has often been considered in opposition
to or in combination with other containment strategies.  Hasegawa et
al.~\cite{Hasegawa} found that quarantine measures outperform the
random and acquaintance preventive vaccination schemes for what
concerns transmission reduction.  The work of Horstmeyer et
al.~\cite{Horstmeyer} suggested instead that a combination of
self-distancing and isolation is particularly effective to contain a
disease.
Through a delay differential equation model,
Heidecke et al.~\cite{Heidecke} found that the efficacy of the
test-trace-isolate-quarantine is limited and requires to be combined
with other enhanced hygienic measures to achieve disease control. They also
warned upon the self-acceleration of disease spread that can be caused
by limited capacities of tracing.

Some works focused specifically on the contact tracing measure,
investigating the role played by different parameters on its efficacy
for the containment of the spread of infectious pathogens. 
Kerr et al. \cite{Kerr} and Burdinski et al. \cite{Burdinski} found
that the efficacy of contact tracing improves with incidence.
In the specific context of
the early COVID-19 pandemic in Seattle, an agent-based model calibrated to demographic, mobility and epidemiological data predicted that the
contact tracing measure would allow the
reopening of society in the absence of massive vaccination coverage while
maintaining epidemic control, if performed strongly, i.e. with high testing and tracing rates, high quarantine compliance, short testing and tracing delays and moderate mask use \cite{Kerr}.  Similarly, Reyna-Lana et
al. \cite{ReynaLara} concluded, by means of a markovian treatment of a SIR
model for the simultaneous contagion processes of infection and
contact tracing, that the combination of case-isolation and
contact tracing is beneficial to the outbreak containment but requires
high adoption of digital contact tracing apps to identify
superspreaders.  Optimal app coverage was also studied by Bianconi et
al. \cite{Bianconi} through a message-passing model. High app adoption,
in particular by high-degree nodes, appeared to be crucial.
However, adoption of digital contact tracing apps seems unlikely to be
achieved by high degree nodes.
Crucially, this was found by Mancastroppa et
al.~\cite{Mancastroppa21} to undermine the performance of digital
contact tracing compared to manual one, as a consequence of a
quenched sampling from the population in contrast with an annealed one.
Through numerical simulations, Hellewell et al. \cite{Hellewelllancet}
and Burdinski et al. \cite{Burdinski} concluded that
efficacy of the forward contact tracing measure is limited.

According to the simulation work of Kojaku et al. \cite{kojaku21} on
synthetic and empirical contact networks, tracing the potential
infector of a known case instead of its potential infectees (backward
instead of forward contact tracing) was exceptionally efficient at
detecting superspreading events, since it leverages two statistical
biases.  Homophily in adoption of digital contact tracing apps
leads to improved performance of contact tracing
when coverage is low \cite{Rizi, Burgio}.
Also the clustering of networks, appears to favor CT performance
in many settings \cite{House, Burdinski, Kiss}.

Despite intense recent activity, an analytical understanding of the
impact of imperfect adherence and implementation delays on the
efficacy of contact tracing is still missing. In this work
we fill this gap, by considering an
SIS-based compartmental model for the self-isolation of symptomatic
individuals and the quarantine of their asymptomatic contacts. More
specifically, we study the influence of three kinds of imperfect
adherence to self-isolation and to quarantine~--~delay to isolation,
imperfect compliance and anticipated exit from isolation~--~on the value
of the epidemic threshold.
Within a mean-field approach, we derive an analytical expression
for the epidemic threshold, defined as the critical virus
transmissibility that separates a healthy absorbing phase from an
endemic phase. This allows to evaluate the performance of
contact tracing and compare it to the efficacy of self-isolation
alone as a function of the parameters describing behavioral and
physiological features of the population.
We further determine analytically the role of
the contact tracing measure on the stationary fractions of infected
individuals.
Finally, we show that heterogeneities in contact patterns and
more complex in-host disease progression do not qualitatively
alter the findings of the mean-field approach.

\section{The Model and its mean-field solution}

The model we consider is a variation of the SIS-based epidemic model
with self-isolation, delay and fatigue developed in Ref.~\cite{QDF}. We
shall hereafter refer to it as "IDF" (isolation-delay-fatigue).

\begin{figure*}
    \centering
    \includegraphics[width=0.9\columnwidth]{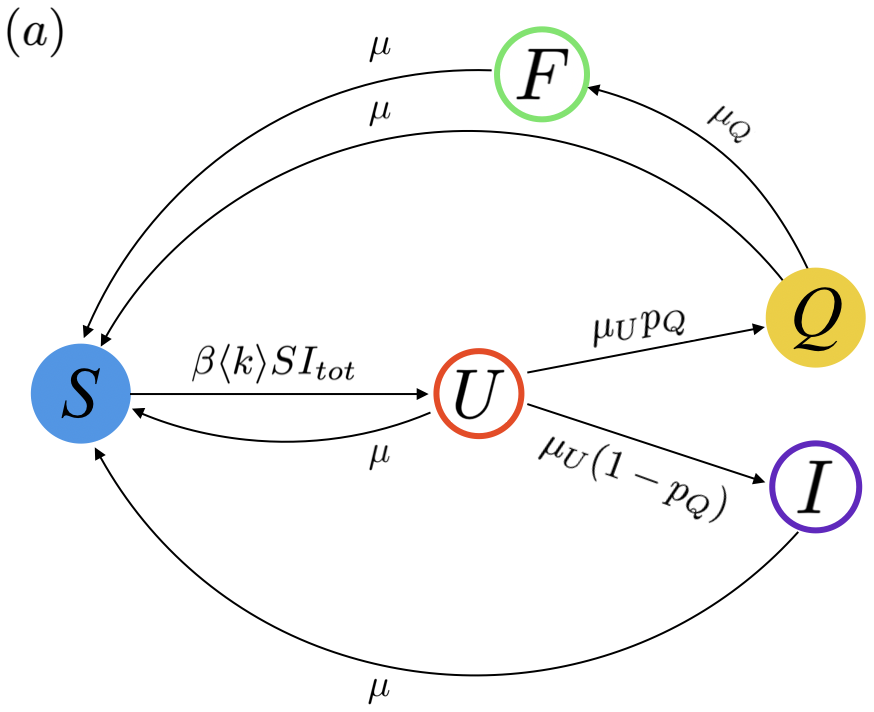}
    \includegraphics[width=0.9\columnwidth]{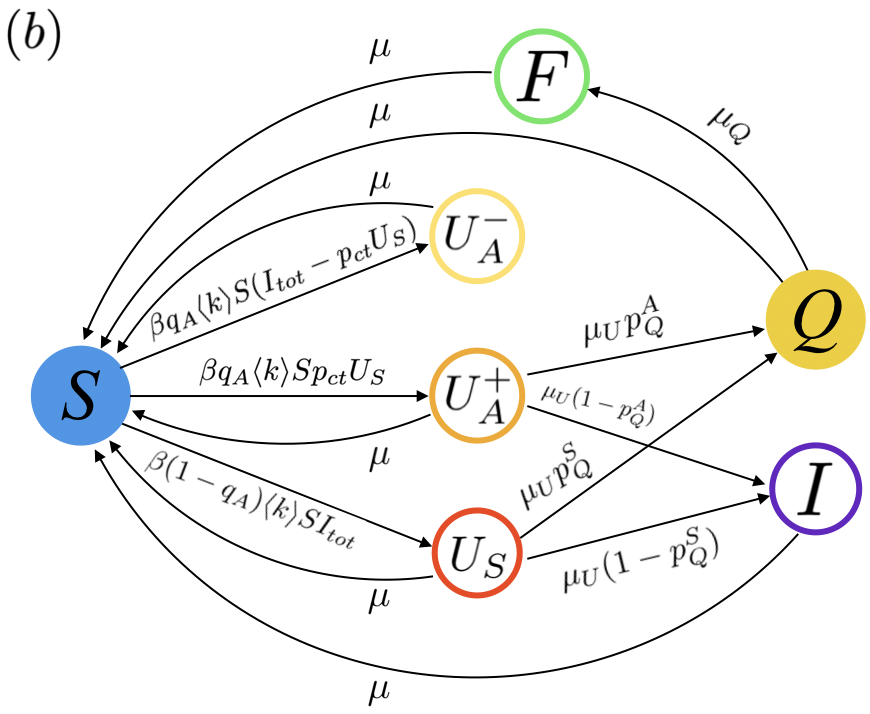}
    \caption{Schematic description of the transitions between
      compartments in (a) the model with self-isolation, delay, fatigue and
      (b) the same model with the addition of contact tracing.
      The quantity $I_{tot}$ is the fraction of individuals
    that are infectious: $I_{tot}=U+I+F$ in (a) and $I_{tot}=U_S+U_A^++U_A^-+I+F$ in (b).}
    \label{fig:scheme-contacttracing-2}
\end{figure*}

According to the IDF dynamics (see Fig.~\ref{fig:scheme-contacttracing-2}(a)),
the contact of a susceptible ($S$)
individual with an infectious one (state $U$, $I$, or $F$, see below)
leads to the infection of the former
with rate $\beta$. Newly infected individuals are assumed to be
immediately infectious but not yet settled on whether to enter
isolation or not (undecided, $U$). After a time interval distributed with
Poissonian rate $\mu_U$ they decide (with probability $p_Q$)
whether to fully interrupt contacts with the rest of the population by
entering the isolated compartment $Q$ or (with probability $1-p_Q$)
to disregard their infectious state
and keep the same rate of interactions with the community, by transitioning
to the $I$ compartment.
The delay between infection and isolation (of mean duration $T_U=1/\mu_U$)
models logistical delays as well as behavioral ones.
In order to account for isolated individuals exiting isolation
before being fully recovered, as a consequence of fatigue, 
a transition from $Q$ to another infectious compartment (fatigued, $F$) occurs
at rate $\mu_Q$.

In the present work, infected individuals are assumed to either stay
asymptomatic through the whole infectious period (with probability
$q_A$) or to develop symptoms (with the complementary probability $1-q_A$).
For this reason the original $U$ compartment is split here into three
compartments.
Individuals developing symptoms enter compartment $U_S$ upon infection.
Being aware of their infected status, symptomatic individuals all go
through the decision process (still with rate $\mu_U$)
on whether to isolate (going to state $Q$, with probability $p_Q^S$)
or not (going to state $I$, with complementary probability $1-p_Q^S$).
Asymptomatic individuals can instead only initiate the decision
process if they are infected by a symptomatic individual (who traces them).
We therefore distinguish
between the asymptomatic individuals who are infected by a
symptomatic individual $U_S$ and are thereby traced ($U_A^+$)
and those who are not ($U_A^-$).
Imperfect tracing capacity is taken into account with
only a fraction $p_{CT}$ of the contacts of symptomatic individuals being successfully traced.

Symptoms are assumed to appear immediately upon infection and thus to
lead to immediate tracing of the contacts of $U_S$ individuals.
Traced asymptomatic individuals ($U_A^+$) decide with rate $\mu_U$
whether to enter the quarantined $Q$ state or not
(thus transitioning to the $I$ state).
As symptoms likely play a major role in determining compliance to
containment measures, we consider the probability
$p_Q^A$ that a traced asymptomatic decides to self-isolate distinct
from (and in particular smaller than or at most equal to) the analogous
probability $p_Q^S$ for a symptomatic individual.
Again, individuals in $Q$ may exit isolation before being fully
recovered by transitioning through compartment $F$, as a consequence
of fatigue.
We assume the same transmissibility across all infectious compartments
$U_A^-$, $U_A^+$, $U_S$, $I$ and $F$ and perfect isolation of
individuals while residing in compartment $Q$.
As the progression of the disease does not depend on the isolation
status, spontaneous recovery transitions may occur from states $U_A^-$,
$U_A^+$, $U_S$, $I$, $Q$ and $F$ to the susceptible state $S$, at the same
recovery rate $\mu$.

For a complete summary of all the transitions and their respective rates
see Appendix~\ref{Model}. The parameters of the model are summarized in Table\ref{tab:parameters}.
Fig.~\ref{fig:scheme-contacttracing-2}(b) presents a complete
description of the epidemic compartments and the transition
rates at the homogeneous mean-field level, where each individual has
$\langle k \rangle$ contacts.

\begin{table}[ht]
\caption{Model parameters}
\begin{tabular}{cc}
    \hline
    \textbf{Parameters} & \textbf{Description}\\
    $\beta$ & rate of infection\\
    $q_A$ & share of asymptomatic\\
    $\mu$ & rate of recovery\\
    $\mu_U$ & rate of decision\\
    $\mu_Q$ & rate of exit from isolation\\
    $p_{CT}$ & probability of being traced\\
    $p_Q^S$ & compliance probability if symptomatic\\
    $p_Q^A$ & compliance probability if asymptomatic\\
    $\langle k \rangle$ & mean node degree
\end{tabular}
\label{tab:parameters}
\end{table}

In this setting the dynamics of the fractions of individuals in states
$U_A^-$, $U_A^+$, $U_S$, $Q$, $I$ and $F$ is governed by the following
set of differential equations

\begin{align}
\begin{cases}
\dot U_A^- &= \beta q_A \langle k \rangle S (U_A^- + U_A^+ + I+F + U_S (1- p_{CT}))+\\
 & \hspace{0.6 \linewidth} - \mu U_A^-\\
\dot U_A^+ &= \beta q_A \langle k \rangle S U_S p_{CT} - (\mu +\mu_U)U_A^+\\
\dot U_S &= \beta (1-q_A) \langle k \rangle S (U_A^- +U_A^+ +I+F+U_S)+\\
& \hspace{0.6 \linewidth} -(\mu+\mu_U)U_S\\
\dot Q &= \mu_U p_Q^A U_A^+ +\mu_U p_Q^S U_S -(\mu+\mu_Q)Q\\
\dot I &= \mu_U (1-p_Q^A) U_A^+ +\mu_U (1-p_Q^S)U_S-\mu I\\
\dot F &= \mu_Q Q - \mu F,
\end{cases}
\end{align}
where the notation $\dot X = \frac{dX}{dt}$ indicates the time
derivatives of the fractions of individuals in each compartment and
$S = 1 - U_A^- - U_A^+ - U_S -Q - I - F$.

The Jacobian matrix obtained by linearization around the disease-free
equilibrium
$(S, U_A^-, U_A^+, U_S, I, Q, F) = (1,0,0,0,0,0,0)$
has 4 real eigenvalues that are always negative and 2 other real
eigenvalues which
become positive as $\lambda=\beta/\mu$ is increased.
The largest one becomes positive (thus making the disease-free equilibrium
unstable) above the epidemic threshold
\begin{equation}
    \lambda_c = \frac{1 - \sqrt{1 - 2 \chi}}{\chi} \cdot \lambda_c^{IDF}(q_A),
    \label{eq:epidemic_threshold}
\end{equation}
where
\begin{align}
  \begin{cases}
    &\lambda_c^{IDF}(q_A) = \frac{1}{\langle k \rangle}
    \frac{1}{1 - \frac{p_Q^S (1-q_A)}{(1 + \frac{T_U}{T})(1 + \frac{T}{T_Q})}} \\
    &T=1/\mu\\
    &T_U=1/\mu_U\\
    &T_Q=1/\mu_Q\\
    &\chi (q_A, p_Q^S, p_Q^A) = \frac{2 q_A (1-q_A) (p_{CT} p_Q^A)}{(1 + \frac{T}{T_U})(1 + \frac{T_U}{T})(1 + \frac{T}{T_Q})} \times\\
    & \hspace{0.3 \linewidth} \times \frac{1}{\left(1 - \frac{p_Q^S (1-q_A)}{(1 + \frac{T_U}{T})(1 + \frac{T}{T_Q})}\right)^2}.
\end{cases}
\end{align}

We observe that the epidemic threshold is given by the expression
for the IDF case (i.e. in the absence of contact tracing, taking into
account that a fraction $1-q_A$ of the individuals is symptomatic and self-isolates)
multiplied
by a factor depending on the various timescales and behavioral
parameters of the model, combined in the single quantity $\chi$.
As shown in Appendix~\ref{lambdareal},
$\chi$ lies in the range between $0$ and $1/2$,
for all values of the parameters, thus
ensuring that the epidemic threshold is always real.

The factor $\frac{1-\sqrt{1-2\chi}}{\chi}$ in
Eq.~\eqref{eq:epidemic_threshold}
is an increasing function of $\chi$ growing from 1 (for $\chi=0$)
to 2 (for $\chi=1/2$).
This leads to the remarkable conclusion that the quarantine of asymptomatic
individuals, possible because of the contact tracing procedure,
leads to an increase of the epidemic threshold that cannot be
larger than a factor 2.
As a consequence, if $\beta/\mu$ for a given pathogen is larger
than twice the critical value $\lambda_c$, contact tracing,
even if perfectly implemented, cannot prevent the epidemic, i.e.,
take the system below the epidemic threshold.

When asymptomatic individuals do not quarantine (because their
compliance probability $p_Q^A = 0$ vanishes or because the share of
traced contacts $p_{CT} = 0$ vanishes), Eq.~\eqref{eq:epidemic_threshold}
gives back the IDF result, $\lambda_c^{IDF}(q_A)$.
Similarly, the IDF result is recovered when all individuals are
either symptomatic ($q_A = 0$) or asymptomatic ($q_A = 1$), trivially
because contact tracing is deactivated by the absence of individuals to be
traced or individuals triggering the tracing, respectively.
In the latter case, when all individuals are asymptomatic,
we recover $\lambda_c^{IDF}(q_A=1) = 1/\langle k \rangle$, the standard SIS result.
Moreover, we notice that in the expression for the
epidemic threshold an imperfect tracing capacity $p_{CT} < 1$ simply
acts as a rescaling of the probability $p_Q^A$ that traced
asymptomatic individuals will quarantine.

Eq.~\eqref{eq:epidemic_threshold} points out that
the epidemic threshold can diverge for perfect contact tracing and perfect compliance to isolation
($p_Q^S \to 1$, $T_U/T \to 0$ and $T_Q/T \to \infty$)
if only symptomatic infections are present ($q_A = 0$).
A diverging threshold means that no pathogen,
no matter its transmissibility $\beta$, can become endemic.
Instead, in a population where a share of individuals
develops asymptomatic forms of the infection ($q_A>0$),
$\lambda_c$ is necessarily finite and there is no way
(even with perfect contact tracing and perfect compliance to
isolation) to eradicate extremely infective pathogens.

\section{Efficacy of contact tracing}

A quantitative measurement of the effect of CT is provided by
the ratio between the epidemic threshold in the case of full
compliance of asymptomatic individuals to isolation and in the case
where they do not isolate at all (similarly to Ref.~\cite{Heidecke}).
It is highly unlikely that asymptomatic individuals are more compliant to the
self-isolation prescription than symptomatic individuals; hence
$\text{max}(p_Q^A) = p_Q^S$.
Indeed, mild symptoms or lack of symptoms
may ruin the motivation to respect isolation, as physical conditions
are not an impediment to carry out the daily routine.
The efficacy of contact tracing can then be defined as
\begin{align}
    \epsilon_{CT} &= \frac{\lambda_c(p_Q^A = p_Q^S)}{\lambda_c(p_Q^A = 0)} \nonumber\\
    &= \frac{1 - \sqrt{1 - 2 \chi(p_Q^A = p_Q^S)}}{\chi(p_Q^A = p_Q^S)}.
    \label{epsilon}
\end{align}

This quantity is bounded in $\epsilon_{CT} \in [1, 2]$: as already discussed above
the contact tracing measure may only bring about a limited
increase of the epidemic threshold.
More dramatic effects on the threshold value may be due to the
$\lambda_c^{IDF}(q_A)$ factor,
i.e., to the self-isolation of symptomatic individuals.

From Eq.~\eqref{epsilon} it is possible to get insight on how virus
characteristics ($q_A$, $\mu$) and behavioral parameters
($p_Q^S$, $\mu_U$, $\mu_Q$) influence the performance of the contact
tracing measure.
\begin{figure}
  \centering
  \includegraphics[width=0.9\columnwidth]{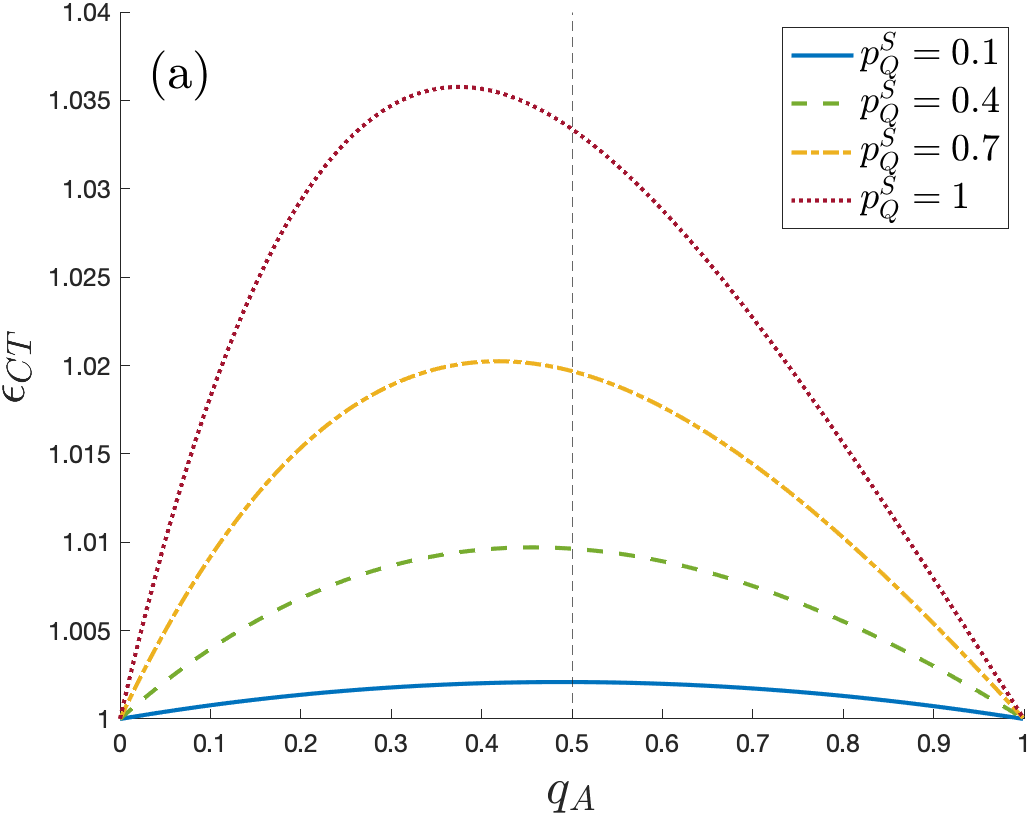}
  \includegraphics[width=0.9\columnwidth]{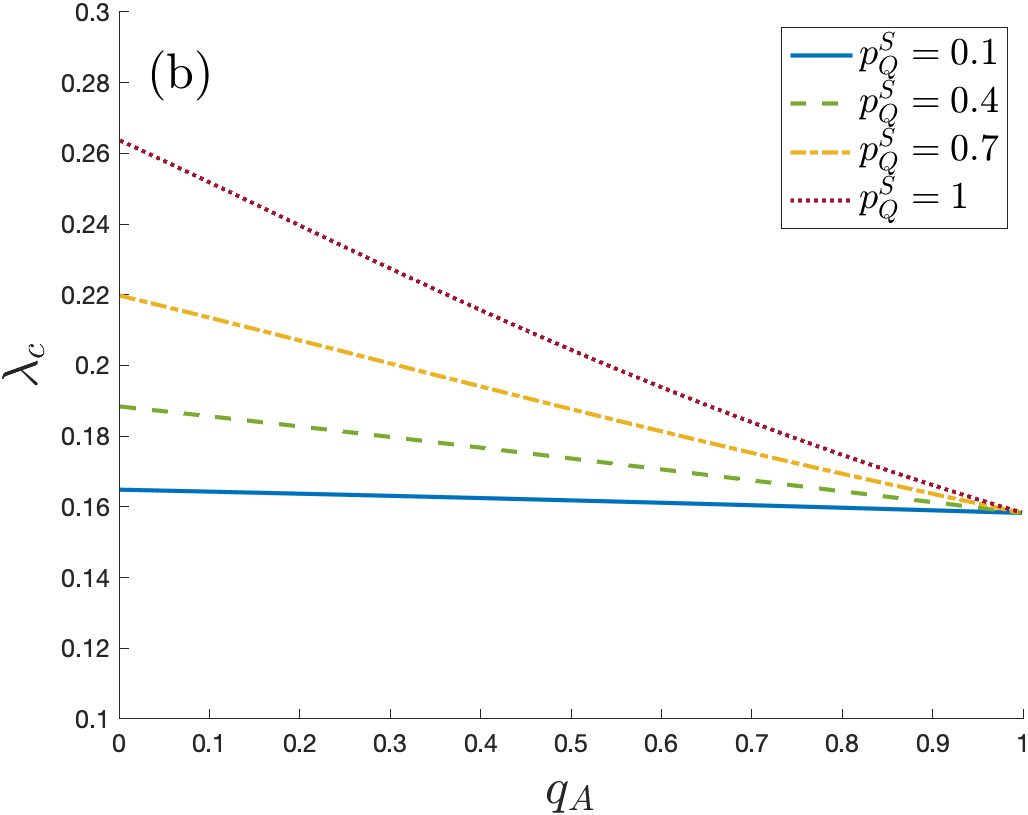}
  \caption{(a) Plot of the efficacy of CT $\epsilon_{CT}$ as a function of $q_A$ for
    various values of $p_Q^S$. (b) Plot of the threshold $\lambda_c$
    against $q_A$ for the same parameters of panel (a). Parameter values: $\mu=\mu_Q=1$, $\mu_U=4$, $p_Q^A=p_Q^S$, $p_{CT}=1$, $\langle k \rangle=6.3$.}
  \label{epsvsq_A}
\end{figure}
For instance, as a function of the share of fully asymptomatic
infections, the efficacy of contact tracing attains a maximum
(see Fig.~\ref{epsvsq_A}(a)) for a value
\begin{align}
    q_A^* &= \frac{(1+T_U/T)(1+T/T_Q) - p_Q^S}{2(1+T_U/T)(1+T/T_Q) - p_Q^S} \nonumber\\
    &= \frac{1}{1+\lambda_c^{IDF}(q_A=0)}.
\end{align}

Interestingly, this value always falls in the range
$q_A^* \in [0, 1/2]$; it is a decreasing function of $p_Q^S$
and of the mean isolation period $T_Q$, while it grows with the delay to isolation $T_U$.
Note however that, while a positive share
of asymptomatic individuals $q_A^* >0$ may maximize the efficacy of
contact tracing, it does not maximize the threshold
(Eq.~\eqref{eq:epidemic_threshold}), which reflects the
combined efficacy of
self-isolation of symptomatic individuals and quarantine of their
asymptomatic contacts. Indeed, under the realistic assumption $p_Q^A
\leq p_Q^S$, the epidemic threshold is always maximized by the complete
absence of asymptomatic infections, $q_A = 0$
(see Fig.~\ref{epsvsq_A}(b)). 

\begin{figure*}
  \centering
  \includegraphics[width=0.9\columnwidth]{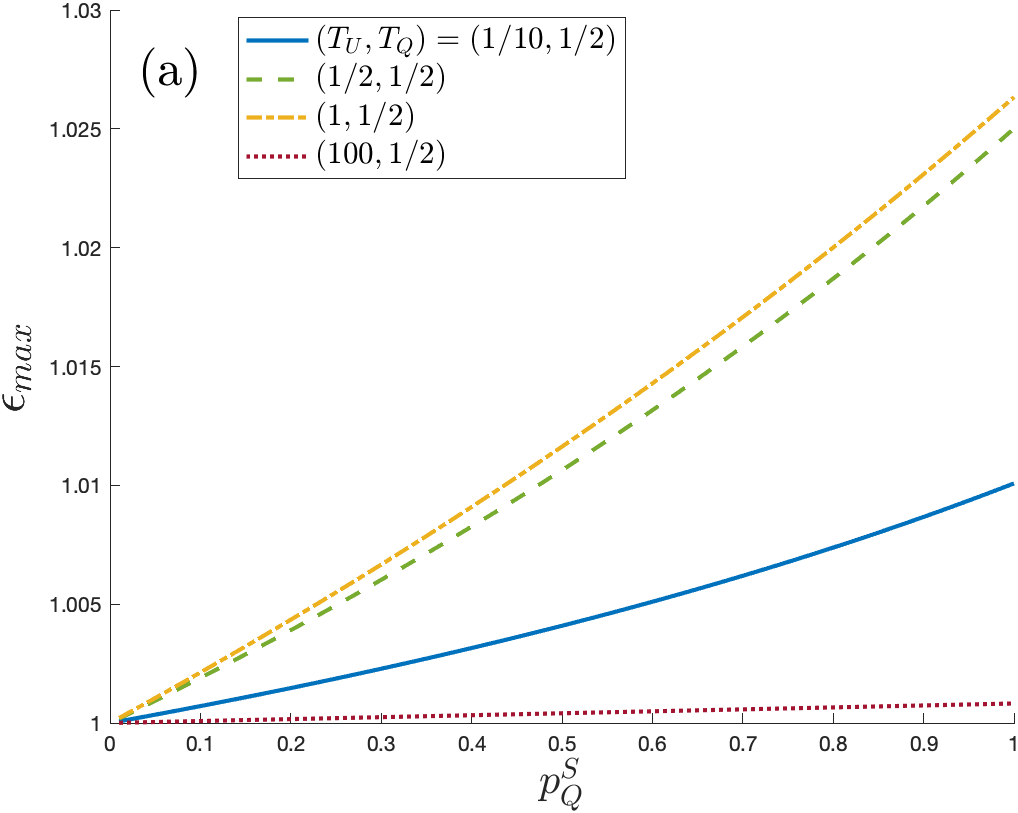}
  \includegraphics[width=0.9\columnwidth]{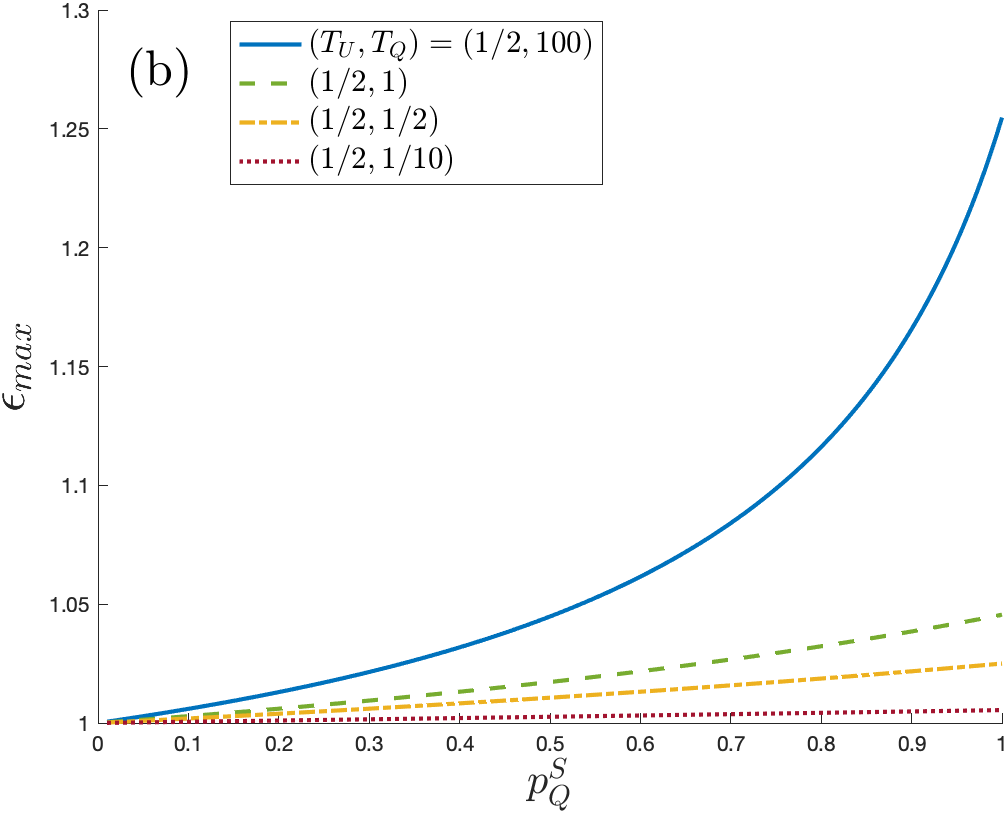}
  \caption{(a) Maximum value of $\epsilon_{CT}$ against $p_Q^S$ for (a) various values of $T_U$ ($T_Q= 1/2$, $T=1$);
    (b) various values of $T_Q$ ($T_U= 1/2$, $T=1$). We consider the case where $p_{CT}=1$.}
  \label{fig:epsilon_max}
\end{figure*}

Plotting $\epsilon_{max}= \epsilon_{CT}(q_A^*)$ as a function of
$p_Q^S$ for various values of $T_U$ and $T_Q$
(Fig.~\ref{fig:epsilon_max}) we find that, despite $\epsilon_{CT}$ assuming
in principle values up to 2, the CT performance is
much more limited: the contribution to the value of the epidemic
threshold due to CT is in practice always of the order of a few percent.

\begin{figure}
  \centering
  \includegraphics[width=0.9\columnwidth]{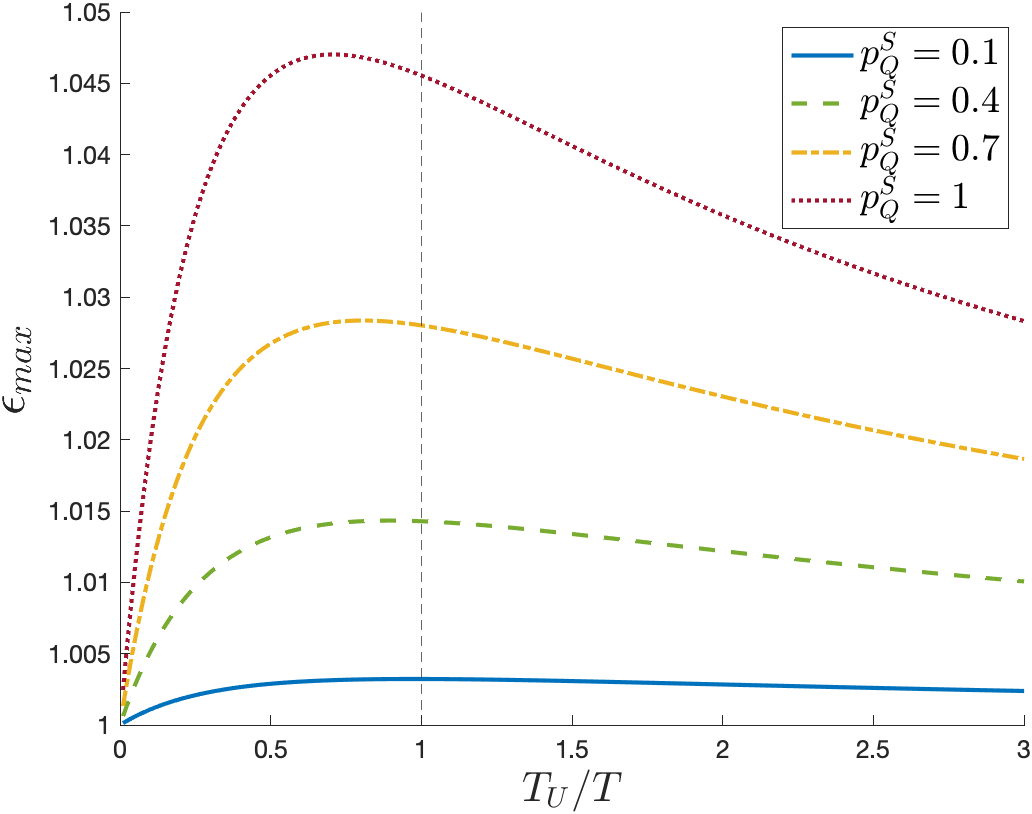}
  \caption{Plot of the maximum value of $\epsilon_{CT}$ against $T_U/T$ for various values
    of $p_Q^S$, showing the presence of a maximum. Parameter values: $T_Q=1$, $p_{CT} = 1$.}
  \label{TUstar}
\end{figure}

Moreover, while $\lambda_c$ is maximised by $T_U \rightarrow 0$, from Eq.~\eqref{epsilon}
we find that the efficacy of CT is
maximized by a delay that can be positive, depending on the value of the other parameters
\begin{equation}
    T_U^* = T \left(1 - \frac{(1-q_A) p_Q^S}{1 + T/T_Q}\right),
\end{equation}
but is nevertheless always shorter than the recovery time $T$ (see
Fig.~\ref{TUstar}).

This is a consequence of a nontrivial tradeoff between two competing
effects.
On the one hand, a short $T_U$ implies that essentially no asymptomatic
is traced.
On the other hand, for very long delays to isolation, many individuals
are traced, but since they take a lot of time to quarantine they may
infect many other individuals, thus reducing the effect of CT.
Assuming that it is possible to act on all parameters, improvement of the efficacy of contact tracing is achieved when $T_U, q_A \rightarrow 0$, $p_Q^A, p_Q^S, p_{CT} \rightarrow 1$, $T_Q \rightarrow \infty$.

In Appendix~\ref{critcompl} we present an analysis of the minimal
values of the compliance $p_Q^S$ or $p_Q^A$ needed to eradicate an
epidemic characterized by a given supercritical transmissibility $\lambda$.

\section{Prevalence in the endemic phase}

This simple model for self-isolation and quarantine allows us also to analytically
compute the stationary fractions of individuals in each
compartment:
\begin{align}
    \label{stationary}
  (U_A^+)^* &= \frac{q_A (1-q_A)}{(1+T/T_U)^2} p_{CT} \lambda_c \langle k \rangle \cdot \frac{\lambda -\lambda_c}{\lambda} \\ \nonumber
    U_S^* &= \frac{(1-q_A)}{1+T/T_U} \cdot \frac{\lambda -\lambda_c}{\lambda}\\ \nonumber
    Q^* &= \frac{1}{\lambda \langle k \rangle}(1-\lambda_c \langle k \rangle + \lambda \langle k \rangle \times \nonumber\\ \nonumber
    &\times \frac{1+a p_Q^S-\sqrt{(1-ap_Q^S)^2 - 4\frac{a}{1+T/T_U} p_{CT} p_Q^A q_A}}{2})\\ \nonumber
    F^* &= \frac{T}{T_Q}Q^*\\ \nonumber
    I^* &= \frac{T}{T_U}(1-p_Q^S)(U_S)^*+\frac{T}{T_U}(1-p_Q^A)(U_A^+)^*\\ \nonumber
    (U_A^-)^* &= (1+T/T_U)[\frac{q_A}{1-q_A} U_S^* - (U_A^+)^*],
\end{align}
where $a = \frac{(1-q_A)}{(1+\frac{T_U}{T})(1 + \frac{T}{T_Q})}$.

In the limit where $q_A \rightarrow 0$, we recover the prevalences
of the IDF model~\cite{QDF}, where the fractions $(U_A^+)^*$ and
$(U_A^-)^*$ vanish, while
$U_S^* \rightarrow \frac{1}{1+T/T_U}\frac{\lambda-\lambda_c}{\lambda}$.
In the same limit, the fraction of isolated and
quarantined individuals reduces to
$Q^* = \frac{p_Q^S}{(1+T/T_Q)(1+T_U/T)} \frac{\lambda-\lambda_c}{\lambda}$.

\begin{figure}
  \centering
  \includegraphics[width=0.9\columnwidth]{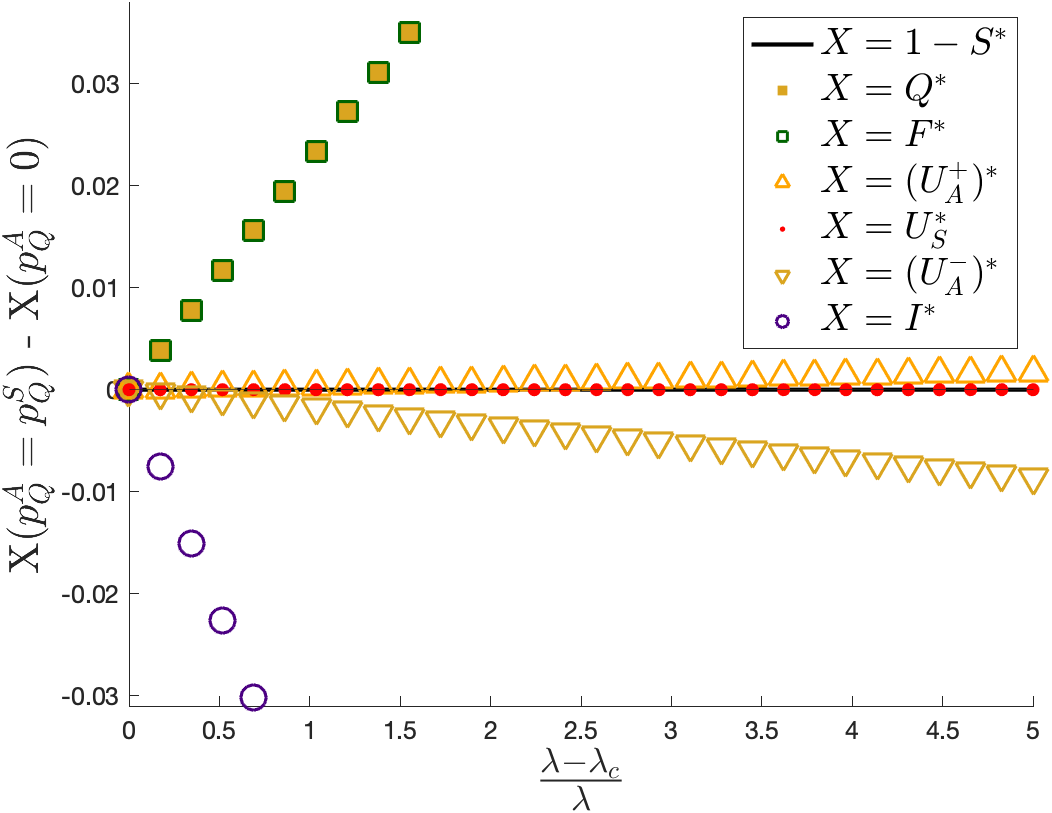} 
  \caption{Plot of the difference between the stationary fractions of individuals
    with ($p_Q^A=p_Q^S=0.9$) and without ($p_Q^A=0$) CT, as a function of the distance
    $(\lambda-\lambda_c)/\lambda$ from the threshold.
    Parameter values: $\mu = 1$, $\mu_U = 4$,
    $\mu_Q = 1$, $q_A = 0.5$, $p_{CT} = 1$, $\langle k \rangle =
    6.3$.}
  \label{fig:prevalence}
\end{figure}

Considering a virus transmissibility standing at a fixed distance
$\frac{\lambda - \lambda_c}{\lambda}$ from the epidemic threshold we
are able to compare the dependence on $p_Q^A$ of these stationary fractions of individuals
above $\lambda_c$.
At a fixed distance from the epidemic threshold,
the stationary fraction of symptomatic individuals $U_S^*$ 
remains unaffected by changes in the compliance $p_Q^A$ of asymptomatic
individuals.  However, since the epidemic threshold is an increasing
function of $p_Q^A$, $(U_A^+)^*$ increases with
$p_Q^A$ while $(U_A^-)^*$ decreases. The quarantine of
asymptomatic individuals therefore enhances the system's tracing
capacity. Having $U_A^+$ individuals transitioning to $Q$ instead of
$I$ indeed reduces the chances of having infectors $I$ that do not
trace their asymptomatic contacts.  The total fraction of undecided
individuals $(U_A^+)^* + (U_A^-)^* + U_S^*$ is expected overall to decrease, while the fractions of individuals in $Q$ and $F$ increase when asymptomatic compliance $p_Q^A$ increases.

This is supported by Fig.~\ref{fig:prevalence} where we see, for a
given choice of the parameter values, the increasing or decreasing
effect of $p_Q^A$ on the quasistationary values, at a given distance
from the epidemic threshold. We consider the stationary fractions of individuals in
a generic compartment $X$ when $p_Q^A = p_Q^S$ (contact tracing is
maximally operative) and when $p_Q^A = 0$ (contact tracing is not
active), as functions of $(\lambda-\lambda_c(p_Q^A=p_Q^S))/\lambda$
and $(\lambda-\lambda_c(p_Q^A=0))/\lambda$, respectively. We then plot
the difference between the stationary fractions of individuals in these limit cases,
so that positive values indicate that the activation of contact
tracing populates the corresponding compartment.  We find that the
amount by which $(U_A^+)^*$ and $I^*$ decrease is perfectly balanced by
the amount by which $Q^*$, $F^*$ and $(U_A^-)^*$ increase, overall
resulting in a fraction of infected individuals $Q^* + I_{tot}^* = 1 -
S^*$ invariant under changes in adherence of asymptomatic individuals,
at a fixed distance from the epidemic threshold.

At a fixed spreading rate, the shift of the epidemic
  threshold induced by the implementation of the contact tracing measure,
  reduces the distance of the system from the critical point, in
  the supercritical regime. This has the effect of making the system
  reach a lower stationary state (and also more slowly) than in the
  absence of contact tracing.

\section{Numerical simulations}

In order to check whether the results obtained in the
mean-field setting also hold for the fully stochastic dynamics,
we perform numerical simulations using a Gillespie
optimized algorithm~\cite{Cota2017} to implement the SIS-like dynamics
on networks built according to the uncorrelated configuration
model~\cite{Catanzaro2005}.
We consider power-law degree-distributed networks with
exponent $\gamma$ and network size $N=10^4$.
The node degrees are constrained in the
range~$k \in [k_{min} = 3, k_{max} = \sqrt{N}]$~-- in order to
have an uncorrelated network without multiple and self connections.
In this setting, we estimate the epidemic threshold by finding the
value of $\lambda$ at which the susceptibility of the system reaches a
maximum~\cite{Ferreira2012}.
We implement the Quasistationary State method (QS)~\cite{Ferreira2012},
for which the dynamics never allows the system to reach the healthy
absorbing state.
We consider both homogeneous ($\gamma=10$) and strongly
heterogeneous ($\gamma=2.5$) networks and compare the results with
the mean-field theory.

\begin{figure}
  \centering
  \includegraphics[width=0.9\columnwidth]{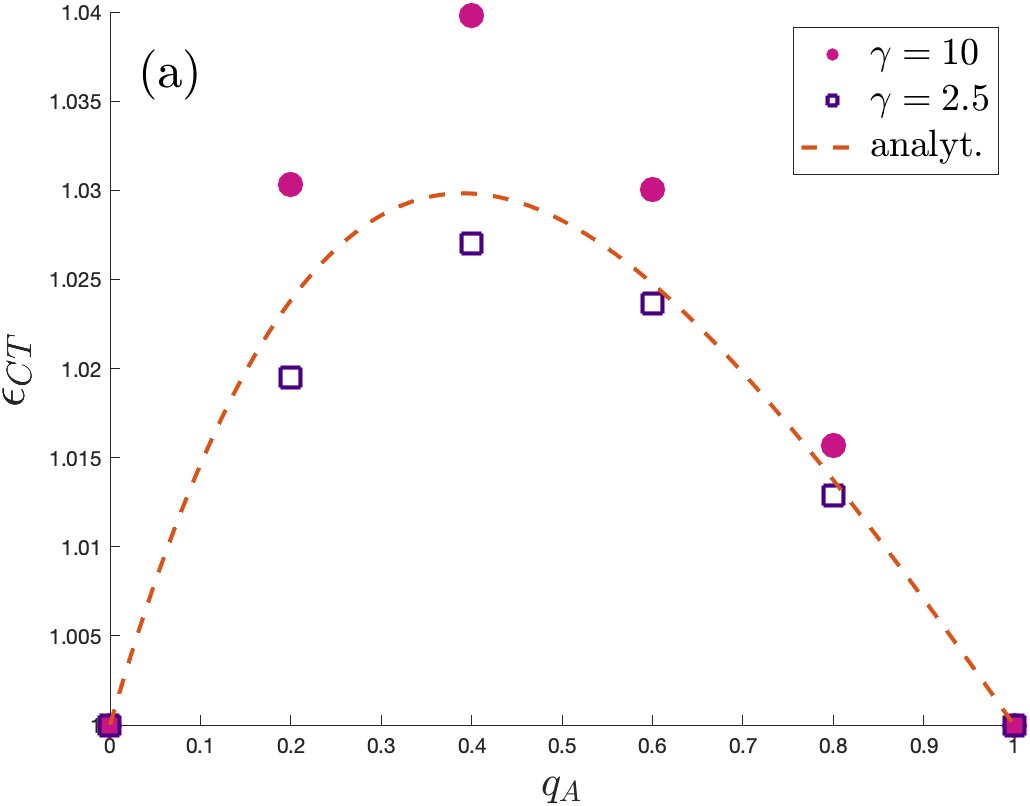}
  \includegraphics[width=0.9\columnwidth]{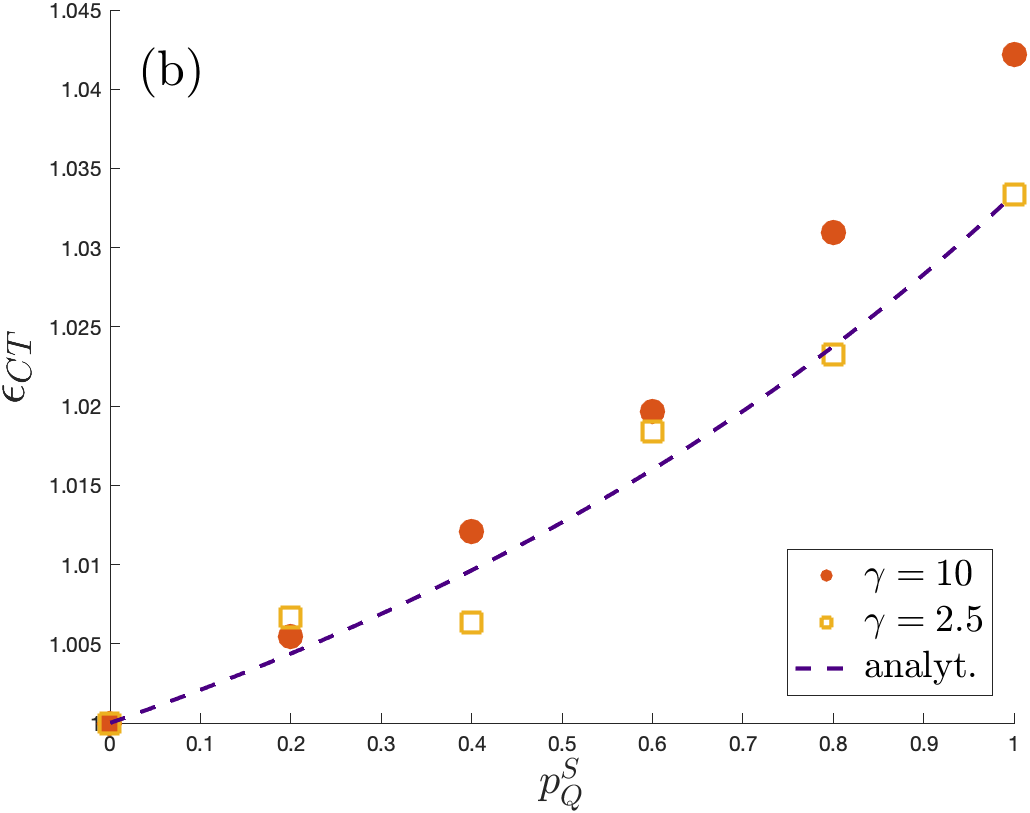}
  \caption{Efficacy of contact tracing in a homogeneous network
    ($\gamma=10$) and in a strongly heterogeneous network
    ($\gamma=2.5$) (analytical prediction, numerical estimates for
    both $\gamma=2.5$ and $\gamma=10$). (a) $\epsilon_{CT}$ as a
    function of $q_A$, with $p_Q^S=0.9$. (b) $\epsilon_{CT}$ as a
    function of $p_Q^S$, with $q_A=0.5$. Parameter values:
    $\mu=\mu_Q=1$, $\mu_U=4$, $p_{CT}=1$.}
  \label{Sim1}
\end{figure}

In Fig.~\ref{Sim1} we show that there is a good agreement between
numerical simulations and analytical results. 

To perform numerical simulations one needs to specify how
the symptomatic/asymptomatic status and the compliance to isolation/quarantine are
chosen for each individual.
To obtain the results presented in Fig.~\ref{Sim1} we have assumed that
the choice is annealed, in the sense
that, for a given individual at each infection event, the development of an asymptomatic form of infection and the decision to isolate are drawn randomly with the corresponding probabilities.
A possibly more realistic alternative is that the probability of
developing an asymptomatic form of infection and the probability of
self-isolating are individual-based. This corresponds to the quenched case, where a given individual always develops the same form of infection (whether symptomatic or asymptomatic) and always takes the same decision concerning isolation (whether it is to isolate or not to isolate).
The immune history of individuals is indeed known to play a role in the
probability of developing asymptomatic forms of infection \cite{Wang2023} while
personal conditions and beliefs concerning self-isolation from the
community likely determine compliance
at an individual level \cite{Smith2020}. 

\begin{figure}
  \centering
  \includegraphics[width=0.9\columnwidth]{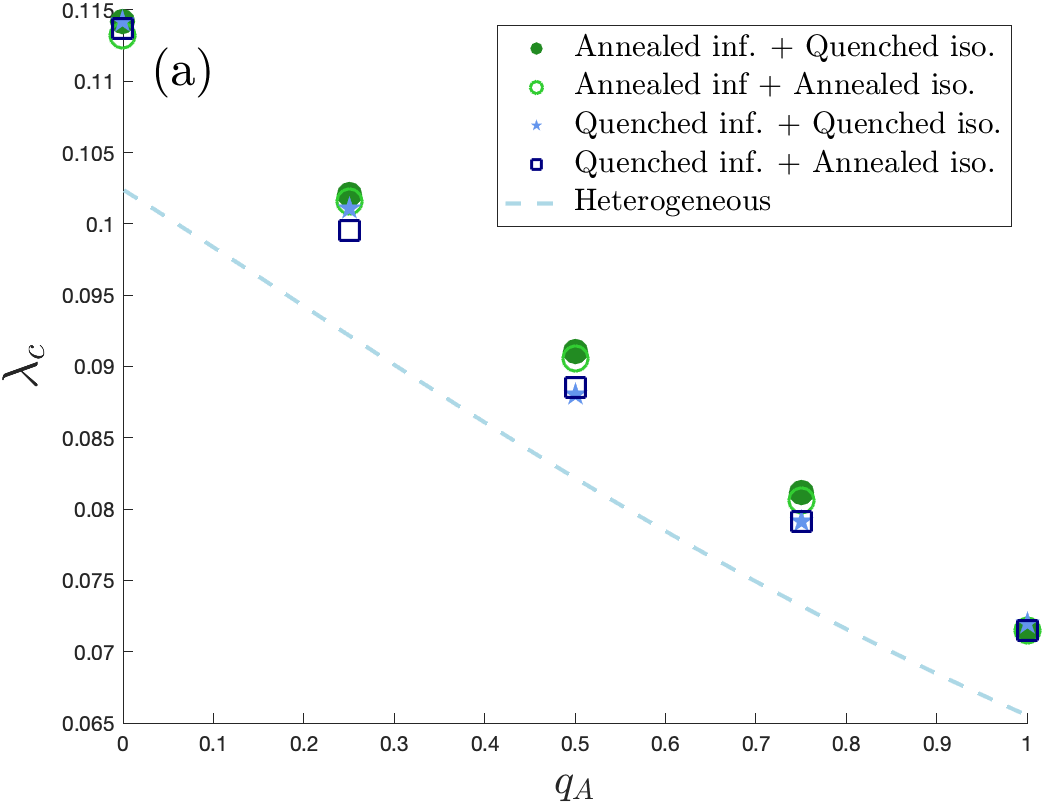}
  \includegraphics[width=0.9\columnwidth]{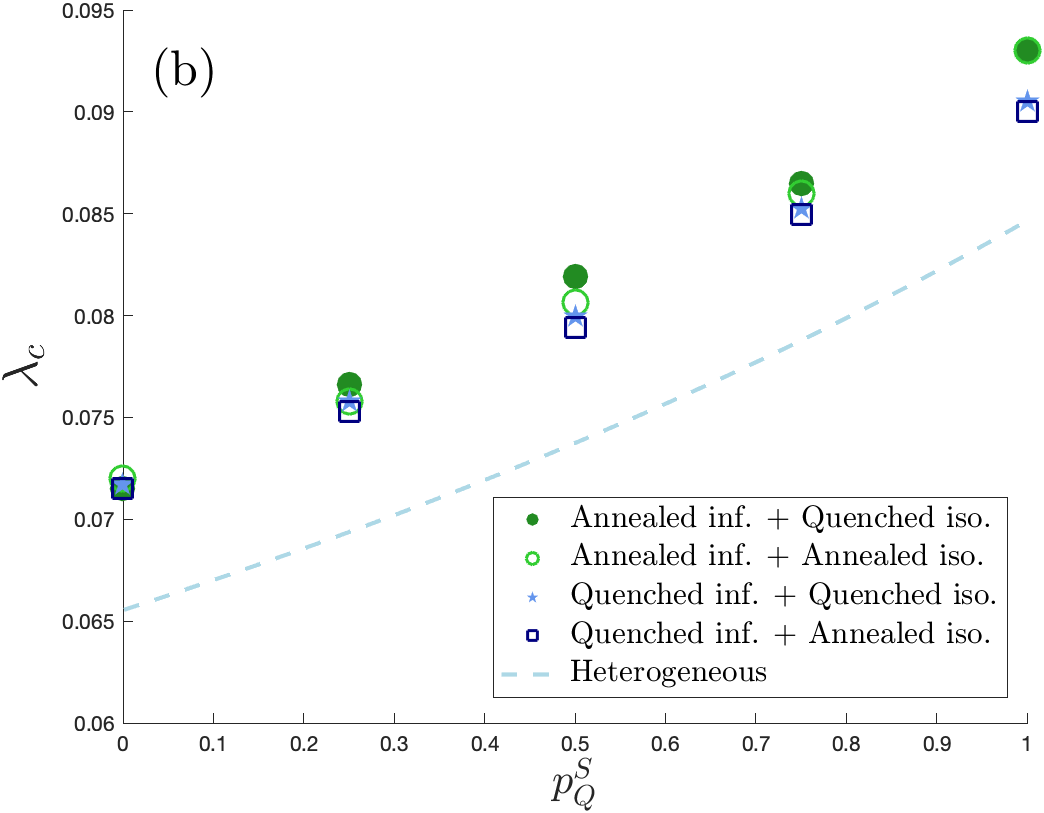}
  \caption{Role of quenchedness (numerical predictions and
    Heterogeneous mean-field estimates achieved by substituting the
    homogeneous mean-field topological factor $1/\langle k \rangle$ by
    $\langle k \rangle/\langle k^2 \rangle$.
    (a) $\lambda_c$ vs $q_A$ for all quenched-annealed combinations, with
    $p_Q^S=0.9$.  (b) $\lambda_c$ vs $p_Q^S$ for all quenched-annealed
    combinations, with $q_A=0.5$. Parameter values: $\mu = \mu_Q = 1$,
    $\mu_U = 4$, $p_{CT} = 1$.}
    \label{Sim2}
\end{figure}

In Fig.~\ref{Sim2} we compare the results we obtain in all possible
scenarios of quenched or annealed treatment of the development of an
asymptomatic form of infection and of the decision of isolating or
not.
The figures show minimal differences between the various cases, even for
a strongly heterogeneous degree distribution.

\section{A more realistic model for In-host Disease Progression}

The stylized model we considered analytically does not only
oversimplify the contact network, it also makes unrealistic
assumptions concerning the way the disease progresses within infected
individuals. In particular, the assumption that infection and recovery
transitions are Poissonian and the constant infectiousness over the
course of infection are unrealistic for the modelling of COVID-19. Moreover,
the latent period during which newly infected individuals are not
infectious yet and the pre-symptomatic period are not considered at
all. These assumptions might strongly affect the efficacy of the
contact tracing and case isolation measures. The model also neglects
the fact that the contact tracing measure is able to isolate not only
asymptomatic contacts but also symptomatic contacts in their
pre-symptomatic phase.

In this section we check whether our conclusions on the efficacy of
the contact tracing measure also hold in a much more complex and
realistic model of disease progression.
We consider the propagation of the Omicron variant of SARS-CoV-2 on
the branching process model for disease propagation introduced in~\cite{lancet}.
Such a model takes into account
different transmissibilities for symptomatic and asymptomatic
individuals, an evolution of infectiousness over the course of the
infection (through a distribution of generation
intervals) with a latent period, and a pre-symptomatic period (through a distribution of
incubation periods). It also models different levels of immunity
across the population (different numbers of doses of vaccine
administered to each individual with a waning of protection against
infection and symptomatic infection informed by available estimates
for the Omicron variant), time varying test sensitivity (sensitivity of antigenic
tests varying over the course of infection) and imperfect adherence to the
measures (delay to isolation, partial contact reduction during the
isolation period, imperfect share of successfully traced contacts,
imperfect compliance to testing and to isolation, anticipated exit
from isolation as a consequence of fatigue).

For a better comparison with our analytical results, we cancel the
effects of heterogeneous immunity across the population (all
individuals are equally susceptible to infection, all individuals
being considered as unvaccinated), the effect of being asymptomatic on
the potential to further transmit the disease (same transmissibility
for symptomatic and asymptomatic individuals) and the role played by
testing (perfect test sensitivity, perfect compliance and vanishing
delay to testing).  We fix our baseline parameters to conditions that
favor the performance of CT: vanishing delay between information of
being infected and isolation, perfect isolation from the community
during the isolation period, long isolation duration (population mean
of 11 days, i.e. 1.6 times the infectious period that we use as a
proxy of the time for recovery), perfect tracing capacity, perfect
compliance to recommendations in terms of isolation duration.

In order to have a quantitative measurement of the efficacy of contact
tracing akin to our definition in the mean-field
model, we take $\epsilon_{CT}$ as the ratio between two
``critical'' basic reproduction numbers, determined in the case
asymptomatic compliance to isolation is maximum ($p_Q^A = p_Q^S$)
and in the case where it is minimum ($p_Q^A = 0$). 
The ``critical'' basic reproduction number for a given set of parameters
is the initial value of $R_0$ that
generates an effective reproduction number $R_{\textrm{eff}} =1$,
when interventions (self-isolation and/or CT) are implemented.

\begin{figure}
    \centering
    \includegraphics[width=8cm]{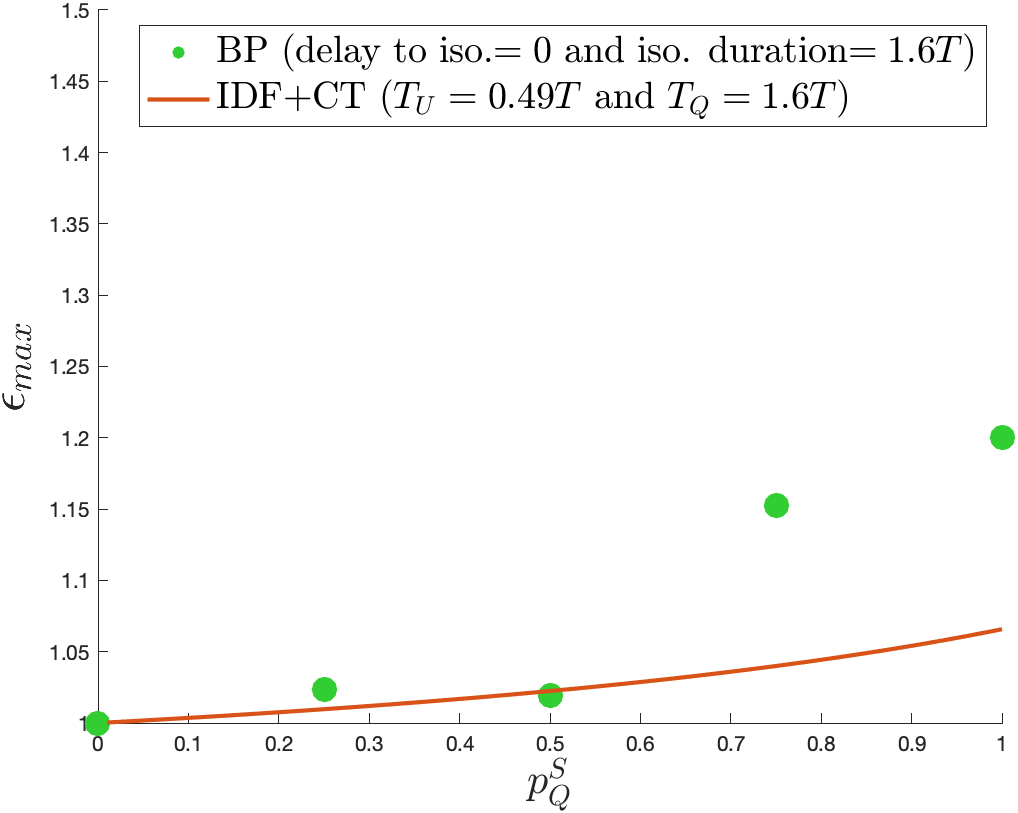}
    \caption{Comparison between estimates of the maximum efficacy
      (as a function of $p_Q^S$) of the
      contact tracing measure in the branching process model of
      \cite{lancet} and in our compartmental model
      (recovery time $T= 1$ and tracing
      capacity $p_{CT} = 1$). ``delay to iso.'' in the BP model refers
      to the delay between information of infection and isolation.}
    \label{fig:bp}
\end{figure}

In Fig.~\ref{fig:bp} we compare the maximum efficacy (as a function of
$q_A$) of the contact tracing measure as computed within our
mean-field model and as estimated through the branching process model.
We approximate the time between infection and information of being
infected with the mean incubation period of the branching process
model. Disregarding the delay between information of being infected
and isolation, this implies a ratio $T_U/T = 0.49$ between the delay
from infection to isolation (approximated by the mean incubation
period of 3.48 days) and the time for recovery (approximated by the
mean infectious period of 7.04 days).  The figure shows that even a
complex and realistic in-host disease progression model predicts a
limited efficacy of the contact tracing measure. Even in extremely
favorable conditions, the increase of the epidemic threshold remains
smaller than $25\%$, in line with our SIS-based compartmental model.

\section{Conclusions}

In this work we developed an SIS-based epidemic model for
self-isolation and (forward and first-order) contact tracing measures
in the presence of imperfect compliance and delays.  We find that the
quarantine of asymptomatic contacts has a very limited impact on
increasing the epidemic threshold. Moreover, it only decreases the
total fraction of infected individuals through the achieved increase of
the epidemic threshold.
It is however especially
useful to quarantine asymptomatic patients in case of outbreaks caused
by viruses that generate low to intermediate shares of asymptomatic
infections, and that propagate in populations where behavioral and
logistical delays to isolation are smaller but close to the time for
recovery.  We find that it is always crucial to incentivate adherence
to isolation, especially for symptomatic individuals.

These conclusions are supported by analytical results in a homogeneous
mean-field setting, allowing for an explicit characterization of the
interplay between disease properties and behavioral
conditions. The role of more realistic contact network characteristics
and more complex in-host disease progression, involving for instance
delayed onset of symptoms is investigated too, with overall
conclusions remaining unaffected.

  In SIR-like models, the
  duration of an outbreak and the height of the peak of the fraction
  of infected individuals are very important observables and the
  efficacy of CT can be measured by the reduction in their values.
  In the present SIS-like model, these observables are not defined.
  The effect of CT on the temporal evolution
  is given here by the increase of the temporal scale $\tau$
  governing the initial exponential growth of the epidemic. Since
  $\tau \propto 1/(\lambda-\lambda_c)$, the effect of CT can be
  ascribed to the increase of the threshold.

In this study we have only considered the effect of forward contact
tracing, as the tracing process allows the quarantine of asymptomatic
individuals who have been in contact with a symptomatic infector.
Backward contact tracing, i.e. the search for the asymptomatic
infector of the symptomatic index case, with the goal of quarantining
him/her and thus preventing further spreading, is known to be highly
effective, in particular in heterogeneous networks~\cite{kojaku21}. It
would be interesting to check how the results presented here are
modified if also backward contact tracing is in place.  A different
model where both types of CT are at work indicates that in such a case
contact tracing may lead to a stronger increase of the epidemic
threshold~\cite{Mancastroppa21}.  Our model also does not study the
tracing of susceptible contacts of index cases (which would represent
an interesting quantity to measure the social weight of the intervention) nor
the tracing of infected contacts that were not infected by the index
case.  A model by Lee et al.~\cite{Lee} focuses on these two aspects
of CT, which however do not change the value of the epidemic threshold.
Indeed, the former does not isolate infected individuals and the latter
necessitates the contact between two infected individuals that becomes
irrelevant near the disease-free equilibrium.

The present study focuses on a population with homogeneous infection rates,
thus neglecting heterogeneities in immune history (whether provided by
previous infections or vaccination) and in age-related immune
response.  Moreover, the possibility of correlations among individuals
with respect to biological and/or behavioral features is disregarded.
The existence of clustering in the contact network has been neglected
as well.  While containment measures such as vaccination and CT are
expected to suffer from assortativity in adherence~\cite{Glaubitz,
  Rizi}, at least in some regimes~\cite{Burgio}, the contact tracing
measure is predicted to benefit from clustering~\cite{House,
  Burdinski, Kiss}.  Finally, we make a set of assumptions that might lead
to overestimates of the efficacy of the contact tracing measure. Our
model neglects the role of diagnostic tests, massively used during the
COVID-19 pandemic. It does not distinguish between the intrinsic
transmissibilities of asymptomatic and symptomatic individuals. Even
symptomatic individuals who do not comply with isolation trace their
contacts, which is not realistic. We moreover assume immediate tracing
of contacts upon infection.

The analysis of modifications of the current framework, where one or
more of these simplyfing assumptions are lifted,
constitutes an interesting avenue for further research.

\section{List of Transitions}
\label{Model}
\begin{itemize}
\item
  Spontaneous decays to $S$
  \begin{eqnarray} \nonumber
   \textrm{Transition} & ~~~~~~ & \textrm{Rate} \\ \nonumber
  I \to S & & \mu\\ \nonumber
  U_S \to S & &\mu\\ \nonumber
  U_A^+ \to S & &\mu\\ \nonumber
  U_A^- \to S & &\mu\\ \nonumber
  Q \to S & &\mu\\ \nonumber
  F \to S & &\mu 
\end{eqnarray}

\item
  Spontaneous decays to $F$
  \begin{eqnarray} \nonumber
    \textrm{Transition} & ~~~~~~ & \textrm{Rate} \\ \nonumber
    Q \to F & &\mu_Q
\end{eqnarray}

\item
  Transitions from Undecided states
  \begin{eqnarray} \nonumber
    \textrm{Transition} & ~~~~~~ & \textrm{Rate} \\ \nonumber
    U_S \to Q & &\mu_U p_Q^S \\ \nonumber
    U_S \to I & &\mu_U (1-p_Q^S) \\ \nonumber
    U_A^+ \to Q & &\mu_U p_Q^A \\ \nonumber
    U_A^+ \to I & &\mu_U (1-p_Q^A)
\end{eqnarray}

\item
  Infections by $U_S$ nodes
  \begin{eqnarray} \nonumber
    \textrm{Transition} & ~~~~~~ & \textrm{Rate} \\ \nonumber
    S+U_S \to U_A^++U_S & &\beta q_A p_{CT} \\ \nonumber
    S+U_S \to U_A^-+U_S & &\beta q_A (1-p_{CT}) \\ \nonumber
    S+U_S \to U_S + U_S & &\beta (1-q_A)
\end{eqnarray}

\item
  Infections by $X=(U_A^+,U_A^-,I,F)$ nodes
  \begin{eqnarray} \nonumber
    \textrm{Transition} & ~~~~~~ & \textrm{Rate} \\ \nonumber
    S+X \to U_A^-+X & &\beta q_A \\ \nonumber
    S+X \to U_S+X & &\beta (1-q_A)
\end{eqnarray}

\end{itemize}

\section{Demonstration that the threshold $\lambda_c$ is always real}
\label{lambdareal}
We want to show that $\chi \le 1/2$. Let us start from the
more restrictive condition $\chi \leq p_{CT}p_Q^A/2$, which can be written
as a second order inequality for $q_A$:
\begin{align}
    q_A^2 &\left[ (p_Q^S)^2 + 4 \frac{T_U}{T T_Q} (T+T_Q) \right] + \nonumber\\
    &q_A \left[ 2 p_Q^S \left(\frac{T}{T_Q} + \frac{T_U}{T} + \frac{T_U}{T_Q}\right) - 4 \frac{T_U}{T T_Q} (T + T_Q)\right] + \nonumber\\
    &\left(\frac{T}{T_Q} + \frac{T_U}{T} + \frac{T_U}{T_Q}\right)^2\geq 0.
    \label{eq:parabola}
\end{align}
We now show that this condition holds for any $q_A\in [0,1]$.

The minimum of the left hand side of Eq.~\eqref{eq:parabola} (L.H.S.)
is always located at values of $q_A < 1$. Indeed,
\begin{align}
    \text{argmin}_{q_A}L.H.S. &< 1 \nonumber\\ -2\frac{T_U}{T T_Q}
    (T+T_Q) - (p_Q^S)^2 &- p_Q^S(\frac{T}{T_Q} + \frac{T_U}{T} +
    \frac{T_U}{T_Q}) < 0,
\end{align}
which is always true.

The condition for the minimum of the L.H.S. to be located
at $q_A>0$ is
\begin{align}
\text{argmin}_{q_A}L.H.S. &>0 \nonumber\\
    p_Q^S &< \frac{2 T_U (T+T_Q)}{T^2 + T_U(T + T_Q)}.
    \label{eq:condition_qA_pos}
\end{align}
Hence it depends on $p_Q^S$ whether the minimum is for $q_A$ smaller
or larger than zero.

In the first case, Eq.~\eqref{eq:parabola} is always satisfied between
$q_A = 0$ and $q_A = 1$, because it is already true for $q_A=0$ and the
L.H.S. is a growing function of $q_A$.

In the other case, Eq.~\eqref{eq:parabola} is always satisfied because
it is satisfied in the minimum. Indeed, the minimum of the L.H.S.
is positive if the following condition on the parameters holds:
\begin{align}
    &\left(\frac{T}{T_Q} + \frac{T_U}{T} + \frac{T_U}{T_Q}\right)^2 \nonumber\\
    \quad \quad & \geq \frac{T_U}{T T_Q}(T+T_Q) - p_Q^S\left(\frac{T}{T_Q} + \frac{T_U}{T} + \frac{T_U}{T_Q}\right).
\end{align}

Under the assumption of $\text{argmin}_{q_A}L.H.S. >0$ (using Eq.~\eqref{eq:condition_qA_pos}), we find that it is always true because:
\begin{align}
    &\left(\frac{T}{T_Q} + \frac{T_U}{T} + \frac{T_U}{T_Q}\right)^2 \geq \nonumber\\
    &>\frac{T_U}{T T_Q}(T+T_Q) - \frac{2 T_U (T+T_Q)}{T^2 + T_U(T + T_Q)}\left(\frac{T}{T_Q} + \frac{T_U}{T} + \frac{T_U}{T_Q}\right)\\
    &= \frac{T_U}{T T_Q}(T+T_Q) - 2 \frac{T_U}{T T_Q} (T+T_Q)\\
    &= - \frac{T_U}{T T_Q}(T+T_Q).\nonumber
\end{align}

We conclude that the condition $\chi \leq p_{CT}p_Q^A/2$ is always met in the
interval $q_A \in [0, 1]$.
When $\text{argmin}_{q_A} L.H.S.< 0$, the
L.H.S. is positive in the interval $q_A \in [0, 1]$ and when
$\text{argmin}_{q_A} L.H.S. > 0$, we have $\text{min}(L.H.S.) > 0$
implying too that $\chi \leq p_{CT}p_Q^A/2$.
Since $p_Q^A,p_{CT}\in[0, 1]$, $\chi\leq1/2$ and the epidemic threshold is real.

\section{Critical compliance}
\label{critcompl}

\begin{figure*}
    \centering
    \includegraphics[width=0.9\columnwidth]{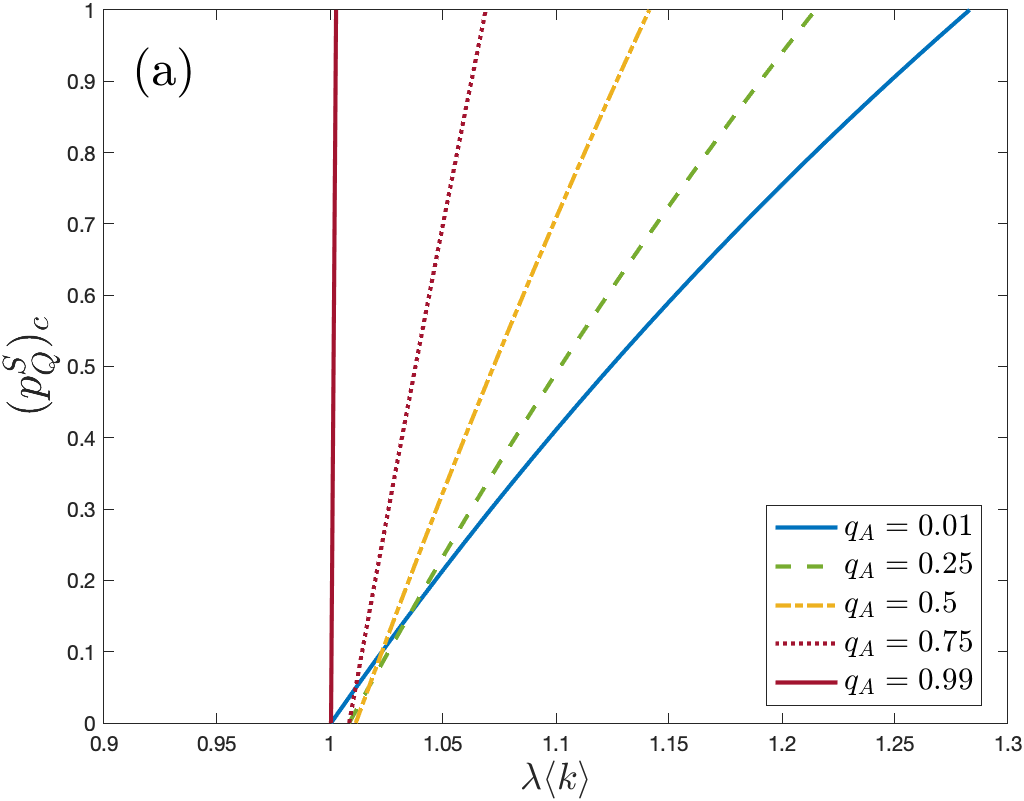}
    \includegraphics[width=0.9\columnwidth]{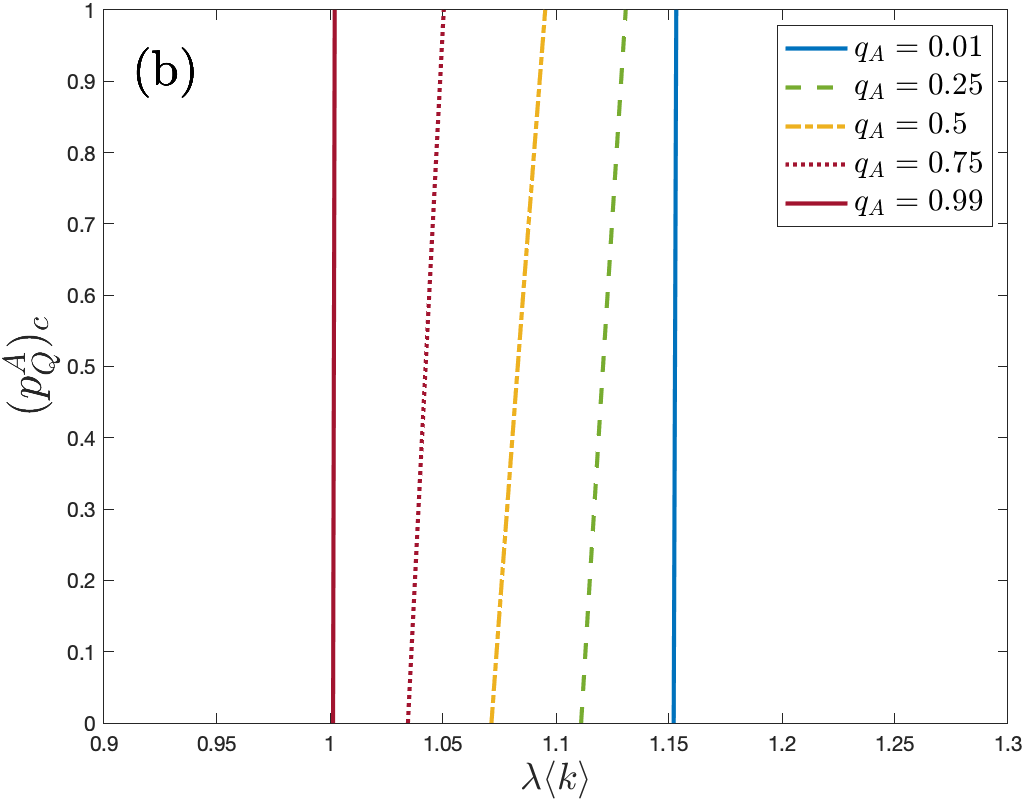}
    \caption{(a) Critical value of the compliance probability of
      symptomatic individuals against $\lambda \langle k \rangle$ for
      various values of asymptomatic probability $q_A \in[0,
        1]$ ($p_Q^A = 0.6$).
(b) Critical value of the compliance probability of
      asymptomatic individuals against $\lambda \langle k \rangle$ for
      various values of asymptomatic probability $q_A \in[0,
        1]$ ($p_Q^S = 0.6$).
      Parameter values: $\mu = 1$, $\mu_U = \mu_Q = 2$.
    }
    \label{fig:pqcrit}
\end{figure*}

In this appendix we study the role of the compliance of symptomatic
and asymptomatic individuals in containing the spread of the epidemic.
Given a pathogen with a specific value of $\lambda$, what are the values
of $p_Q^S$ that are sufficient to make $\lambda<\lambda_c$ so that the epidemic
becomes subcritical and disappears?

Setting $\lambda=\lambda_c$ and inverting Eq.~\eqref{eq:epidemic_threshold}, the expression of $p_Q^S$ that allows the system to reach the epidemic threshold is easily derived 
\begin{equation}
    (p_Q^S)_{c} = \left[1 - \frac{\chi (\lambda / \lambda_c^{IDF}(q_A))^2 + 2}{2 (\lambda \langle k \rangle)}\right]\frac{(1 + T_U/T)(1+T/T_Q)}{1-q_A}.
\end{equation}

This expression for $(p_Q^S)_c$ is not a priori limited
to the $[0, 1]$ interval. Values of $(p_Q^S)_c < 0$ imply that the
system does not require isolation of symptomatic individuals to reach
the epidemic threshold. In practice this means that $\lambda$ is already subcritical in the absence of isolation of symptomatic individuals.
Values $(p_Q^S)_c > 1$ mean instead that even the
isolation of all symptomatic individuals is not sufficient to drive the
system to the disease-free state.

These results can be translated into ranges of values of the
virus transmissibility $\lambda \langle k \rangle$ that the
isolation of symptomatic individuals is able to contain.  Isolation of
symptomatic individuals is not necessary when
\begin{align}
  \lambda \langle k \rangle &\leq \lambda_c^{IDF+CT}(q_A)|_{p_Q^S=0} \langle k \rangle\\
    &= \frac{1- \sqrt{1- 2 \chi}}{\chi},
\end{align}
as the outbreak would be contained regardless of their isolation.  On
the other hand, thanks to the isolation of symptomatic individuals,
the outbreak can be contained for values of the transmissibility rate
$\lambda \langle k \rangle$ larger than the SIS result but as long as
\begin{align}
    \lambda \langle k \rangle &\leq \lambda_c^{IDF+CT}(q_A)|_{p_Q^S=1} \langle k \rangle\\
    &= \frac{1}{1-a} \cdot \frac{1- \sqrt{1- 2 \chi}}{\chi},
\end{align}
where $a = \frac{(1-q_A)}{(1+\frac{T_U}{T})(1 + \frac{T}{T_Q})}$.

In Fig.~\ref{fig:pqcrit}(a) we plot the dependence of $(p_Q^S)_c$ on
$\lambda \langle k \rangle$ for various values of the probability
$q_A$ of being asymptomatic. We see that the isolation of symptomatic
individuals is the more useful to contain the spread of more
transmissible infectious pathogens the lower the share of
asymptomatic individuals in the infected population.

We repeat the above reasoning with the compliance $p_Q^A$ of
asymptomatic individuals with isolation mandates.  The critical value
of $p_Q^A$ that allows to reach the epidemic threshold is
\begin{align}
        (p_Q^A)_c &= \frac{1}{\delta}\frac{1}{\lambda \langle k
    \rangle} \left[ \frac{1}{\lambda^{IDF}(q_A) \langle k \rangle} -
    \frac{1}{\lambda \langle k \rangle} \right],
\end{align}
where
$\delta = \frac{q_A(1-q_A) p_{CT}}{(1+\frac{T}{T_U})(1+\frac{T_U}{T})(1+\frac{T}{T_Q})}$.

The outbreak can be contained by quarantining traced asymptomatics if

\begin{align}
    \lambda_c^{IDF+CT}(q_A)|_{p_Q^A=0} \leq \lambda \leq \lambda_c^{IDF+CT}(q_A)|_{p_Q^A=1}\\
    \frac{1}{1-a p_Q^S} \leq \lambda \leq \frac{1}{1-a p_Q^S} \frac{1-\sqrt{1-2 \chi(p_Q^A=1)}}{\chi(p_Q^A=1)}.
    \label{eq:conditions}
\end{align}

In Fig.~\ref{fig:pqcrit}(b) we show the dependence of the critical value
of $p_Q^A$ on $\lambda \langle k \rangle$ for various values of $q_A
\in [0,1]$. A comparison with panel (a) of the same figure shows that
for the chosen values of the parameters, the
role played by the isolation of symptomatic individuals is
significantly larger than the one played by the quarantine of
asymptomatic individuals.

\bibliography{CCGDM.bib}
\end{document}